# Coupled integral equations method with open boundary conditions for calculation the characteristics of structured waveguides

M.I. Ayzatsky

National Science Center Kharkiv Institute of Physics and Technology (NSC KIPT), 61108, Kharkiv, Ukraine
E-mail: mykola.aizatsky@gmail.com, aizatsky@kipt.kharkov.ua

The results of modification of the CASCIE code aimed at implementing open boundary conditions are presented. The accelerator section developed at CERN was chosen as a prototype for the structured waveguide under testing. Results of testing the CASCIE-M code confirms that the implementation of matrix open boundary conditions gives possibility to consider the structure in which waves enter and exit without additional reflections from couplers.

It was shown that the dependence of the reflection coefficient on frequency differs from the similar dependence for a waveguide with couplers. It does not have a regular sequence of minimum and maximum values associated with reflections from the couplers and the formation of resonance conditions. This indicates that the reflections are of a different nature and are associated with inhomogeneity. The proposed modification of the coupled integral equation method allows us to investigate the accuracy of the field expansion on which coupled mode theory can be constructed that describes structured waveguides.

## 1. INRODUCTION

The most commonly used types of slow-wave electrodynamic systems are structured waveguides[1]. One of the effective approaches for calculating the characteristics of structured waveguides is the coupled integral equations method (CIEM) [1,2,3,4,5,6]. Using this method, it was shown that a structured waveguide can be described by the system of coupled matrix equations. Real electrodynamic systems include a structured waveguide with a finite number of cells. The first and last cells in this case play the roles of couplers with external devices.

However, in some cases it is necessary to investigate the characteristics of a finite structured waveguide without taking into account the influence of the boundaries and the reflections which they produce. For example, such approach is useful under investigating the frequency and size dependencies of the characteristics of structured waveguide characteristics. To do this, it is needed to formulate open boundary conditions that allow waves to enter and exit the finite structured waveguide without additional boundary reflections. In this case there is no need to tune couplers each time we change a frequency or sizes.

New generalization of the theory of coupled modes, proposed in [7,8,9,10], gives possibility to describe a structured waveguide by using ordinary differential equations. Based on a set of eigen waves of a homogeneous periodic waveguide, a new basis of vector functions is introduced that takes into account the non-periodicity of the waveguide. Representing the total field as the sum of these functions with unknown scalar coefficients, a system of coupled equations that determines the dependence of these coefficients on the longitudinal coordinate can be obtained. In some cases, we can reduce this system and use the single mode approximation. Studying of such possibility and accuracy is an important step in the describing of structured waveguides by differential equations. Such an investigation can be carried out by comparing the results obtained using modified coupled mode theory with those obtained using more general codes that can support open boundary conditions. One of such code could be the code similar to CASCIE [6], modified to work with open boundary conditions.

We have included open boundary conditions in the code CASCIE and this paper presents the results of its testing and use for calculating the characteristics of a structured waveguide.

## 2. BASIC EQUATIONS

We will consider only axially symmetric fields with $E_z, E_r, H_\varphi$ components (TM). Time dependence is $\exp(-i\omega t)$.

It can be shown that a structured waveguide can be described by the system of coupled matrix equations
$$\underline{\underline{T}}^{(k)} \underline{Q}^{(k)} = \underline{\underline{T}}^{+(k)} \underline{Q}^{(k+1)} + \underline{\underline{T}}^{-(k)} \underline{Q}^{(k-1)}, \qquad (1)$$
where vectors $\underline{Q}^{(k)}$ are the vector of expansion coefficients of the longitudinal electric field in some cross-section of the $k$- cell, $\underline{\underline{T}}^{(k)}, \underline{\underline{T}}^{+(k)}, \underline{\underline{T}}^{-(k)}$ - square matrices [4,5]. Real electrodynamic systems include a finite structured waveguide with a finite number of cells. In this case we need to formulate additional conditions for the first and last cells

---
[1] Waveguides that consist of similar, but not always identical, cells ( disk-loaded waveguides, chains of coupled resonators, etc.).

$$\underline{\underline{T}}^{(Q_1)}\underline{Q}^{(1)} + \underline{\underline{T}}^{(Q_2)}\underline{Q}^{(2)} = \underline{Z}^{Q(1)},$$
$$\underline{\underline{T}}^{(Q_{N-1})}\underline{Q}^{(N-1)} + \underline{\underline{T}}^{(Q_N)}\underline{Q}^{(N)} = \underline{0}, \quad (2)$$

where the vector $\underline{Z}^{Q(1)}$ describes the external source.

We will formulate open boundary conditions for equations (1) that allow waves to enter and exit the finite structured waveguide without reflection.

Let us consider the infinite structured waveguide which cells are numbered with index $k$: $k = -\infty, ..., +\infty$. Assume that cells with $k \leq N_H$ are identical, cells with $k \geq N - N_H$ are identical too, but have different sizes. The structured waveguide under consideration will be numbered with indexes $k = N_H + 1, ..., N - N_H - 1$. Then we can write such equations

$$\underline{\underline{T}}^{(B)}\underline{Q}^{(k)} = \underline{\underline{T}}^{+(B)}\underline{Q}^{(k+1)} + \underline{\underline{T}}^{-(B)}\underline{Q}^{(k-1)},\ k \leq N_H,$$
$$\underline{\underline{T}}^{(E)}\underline{Q}^{(k)} = \underline{\underline{T}}^{+(E)}\underline{Q}^{(k+1)} + \underline{\underline{T}}^{-(E)}\underline{Q}^{(k-1)},\ k \geq N - N_H, \quad (3)$$

where $\underline{\underline{T}}^{+(B)} = \underline{\underline{T}}^{-(B)}$ and $\underline{\underline{T}}^{+(E)} = \underline{\underline{T}}^{-(E)}$.

Equations (3) have such solutions [4]

$$\underline{Q}^{(k)} = \sum_s \left(\lambda_s^{(B,+)}\right)^{k-1} V_s^{(B)} \underline{U}_s^{(B)} + \sum_s \left(\lambda_s^{(B,-)}\right)^{k-1} R_s^{(B)} \underline{U}_s^{(B)},\ k \leq N_H,$$
$$\underline{Q}^{(k)} = \sum_s \left(\lambda_s^{(E,+)}\right)^{k-N} V_s^{(E)} \underline{U}_s^{(E)} + \sum_s \left(\lambda_s^{(E,-)}\right)^{k-N} R_s^{(E)} \underline{U}_s^{(E)},\ k \geq N - N_H, \quad (4)$$

where $\underline{U}_s^{(B,E)}$ are the eigen vectors of the matrix $\left(\underline{\underline{T}}^{+(B,E)-1}\underline{\underline{T}}^{(B,E)}\right)$, $V_s^{(B,E)}$, $R_s^{(B,E)}$ are the amplitudes of the eigen vectors (complex numbers). If $\theta_s^{(B,E)}$ are the eigen values of the matrix $\left(\underline{\underline{T}}^{+(B,E)-1}\underline{\underline{T}}^{(B,E)}\right)$

$$\left(\underline{\underline{T}}^{+(B,E)-1}\underline{\underline{T}}^{(B,E)}\right)\underline{U}_s^{(B,E)} = \theta_s^{(B,E)} \underline{U}_s^{(B,E)}, \quad (5)$$

then the Floquet coefficients $\lambda_s^{(B,E,+)}$, $\lambda_s^{(B,E,-)}$ are the solutions of the quadratic characteristic equation [4]

$$\lambda_s^{(B,E)2} - \theta_s^{(B,E)}\lambda_s^{(B,E)} + 1 = 0,$$
$$\lambda_s^{(B,E,\pm)} = \theta_s^{(B,E)}/2 \pm \sqrt{\left(\theta_s^{(B,E)}/2\right)^2 - 1}. \quad (6)$$

We will suppose that $\left|\lambda_s^{(B,E,+)}\right| < 1$ and ($\left|\lambda_s^{(B,E,-)}\right| > 1$).

We also suppose that in the left semi-infinite waveguide only one eigen wave ($s = s_0$) with frequency $\omega$ lying in the first pass band propagates towards the considered structured waveguide. Then, using the boundary conditions at infinite we should choose $V_s^{(B)} = 0$, $s \neq s_0$, $R_s^{(E)} = 0$, $s \neq s_0$. $V_{s_0}^{(B)}$ is the amplitude of the incoming wave, $R_{s_0}^{(B)}/V_{s_0}^{(B)}$ and $V_{s_0}^{(E)}/V_{s_0}^{(B)}$ are the reflection and transmission coefficients[2]. Takin it into account, we can write

$$\underline{Q}^{(0)} = \left(\lambda_{s_0}^{(B,+)}\right)^{-1} V_{s_0}^{(B)}\underline{U}_{s_0}^{(B)} + \sum_s \left(\lambda_s^{(B,-)}\right)^{-1} R_s^{(B)}\underline{U}_s^{(B)} = \left(\lambda_{s_0}^{(B,+)}\right)^{-1} V_{s_0}^{(B)}\underline{U}_{s_0}^{(B)} + \sum_s \lambda_s^{(B,+)} R_s^{(B)}\underline{U}_s^{(B)},$$
$$\underline{Q}^{(1)} = V_{s_0}^{(B)}\underline{U}_{s_0}^{(B)} + \sum_s R_s^{(B)}\underline{U}_s^{(B)},$$
$$\underline{Q}^{(N)} = \sum_s V_s^{(E)}\underline{U}_s^{(E)}, \quad (7)$$
$$\underline{Q}^{(N+1)} = \sum_s \lambda_s^{(E,+)} V_s^{(E)}\underline{U}_s^{(E)}.$$

Substituting these expressions into the equations (3) with indexes $k = 1, 2, N-1, N$

$$\underline{\underline{T}}^{(B)}\underline{Q}^{(1)} = \underline{\underline{T}}^{+(B)}\underline{Q}^{(2)} + \underline{\underline{T}}^{-(B)}\underline{Q}^{(0)},$$
$$\underline{\underline{T}}^{(B)}\underline{Q}^{(2)} = \underline{\underline{T}}^{+(B)}\underline{Q}^{(3)} + \underline{\underline{T}}^{-(B)}\underline{Q}^{(1)},$$
$$\underline{\underline{T}}^{(E)}\underline{Q}^{(N-1)} = \underline{\underline{T}}^{+(E)}\underline{Q}^{(N)} + \underline{\underline{T}}^{-(E)}\underline{Q}^{(N-2)}, \quad (8)$$
$$\underline{\underline{T}}^{(E)}\underline{Q}^{(N)} = \underline{\underline{T}}^{+(E)}\underline{Q}^{(N+1)} + \underline{\underline{T}}^{-(E)}\underline{Q}^{(N-1)},$$

we get

---

[2] It must be taken into account that the output field amplitude changes due to attenuation and non-uniform geometry

$$\sum_s R_s^{(B)} \left( \underline{\underline{T}}^{(B)} - \lambda_s^{(B,+)} \underline{\underline{T}}^{-(B)} \right) \underline{U}_s^{(B)} - \underline{\underline{T}}^{+(B)} \underline{Q}^{(2)} = V_{s_0}^{(B)} \left\{ \left( \lambda_{s_0}^{(B,+)} \right)^{-1} \underline{\underline{T}}^{-(B)} - \underline{\underline{T}}^{(B)} \right\} \underline{U}_{s_0}^{(B)},$$

$$\sum_s R_s^{(B)} \underline{\underline{T}}^{-(B)} \underline{U}_s^{(B)} - \underline{\underline{T}}^{(B)} \underline{Q}^{(2)} + \underline{\underline{T}}^{+(B)} \underline{Q}^{(3)} = -V_{s_0}^{(B)} \underline{\underline{T}}^{-(B)} \underline{U}_{s_0}^{(B)},$$

$$\underline{\underline{T}}^{-(E)} \underline{Q}^{(N-2)} - \underline{\underline{T}}^{(E)} \underline{Q}^{(N-1)} + \sum_s V_s^{(E)} \underline{\underline{T}}^{+(E)} \underline{U}_s^{(E)} = 0,$$

$$\underline{\underline{T}}^{-(E)} \underline{Q}^{(N-1)} + \sum_s V_s^{(E)} \left\{ \lambda_s^{(E,+)} \underline{\underline{T}}^{+(E)} - \underline{\underline{T}}^{(E)} \right\} \underline{U}_s^{(E)} = 0.$$

(9)

It can be shown that the sum $\sum_s A_s \underline{U}_s = \underline{\underline{U}} \, \underline{A}$, where $\underline{U}_s$ are the column vectors and $A_s$ - numbers, $\underline{\underline{U}} = \{\underline{U}_s\}$ - a matrix which s-column is the column vectors $\underline{U}_s$ and $\underline{A}$ -a column vector formed from the numbers $A_s$.

Using it and equality $\left\{ \left( \lambda_{s_0}^{(B,+)} \right)^{-1} \underline{\underline{T}}^{-(B)} - \underline{\underline{T}}^{(B)} \right\} \underline{U}_{s_0}^{(B)} = -\lambda_{s_0}^{(B,+)} \underline{\underline{T}}^{-(B)} \underline{U}_{s_0}^{(B)}$ we can rewrite the equations (9) as

$$\underline{\underline{G}}^{(B,1)} \underline{R} - \underline{\underline{T}}^{+(B)} \underline{Q}^{(2)} = -\lambda_{s_0}^{(B,+)} V_{s_0}^{(B)} \underline{\underline{T}}^{-(B)} \underline{U}_{s_0}^{(B)},$$

$$\underline{\underline{G}}^{(B,2)} \underline{R} - \underline{\underline{T}}^{(B)} \underline{Q}^{(2)} + \underline{\underline{T}}^{+(B)} \underline{Q}^{(3)} = -V_{s_0}^{(B)} \underline{\underline{T}}^{-(B)} \underline{U}_{s_0}^{(B)},$$

$$\underline{\underline{T}}^{-(E)} \underline{Q}^{(N_{REZ}-2)} - \underline{\underline{T}}^{(E)} \underline{Q}^{(N_{REZ}-1)} + \underline{\underline{G}}^{(E,2)} \underline{V}^{(E)} = 0,$$

$$\underline{\underline{T}}^{-(E)} \underline{Q}^{(N_{REZ}-1)} + \underline{\underline{G}}^{(E,1)} \underline{V}^{(E)} = 0,$$

(10)

where $\underline{R} = \{R_s\}$ - reflection vector, $\underline{V} = \{V_s\}$ - transmission vector.

Columns of matrix $\underline{\underline{G}}^{(B,1)}$ are vectors $\left( \underline{\underline{T}}^{(B)} - \lambda_s^{(B,+)} \underline{\underline{T}}^{-(B)} \right) \underline{U}_s^{(B)}$; $\underline{\underline{G}}^{(B,2)} = \underline{\underline{T}}^{-(B)} \underline{\underline{U}}^{(B)}$; columns of matrix $\underline{\underline{G}}^{(E,1)}$ are vectors $\left( \lambda_s^{(E,+)} \underline{\underline{T}}^{-(E)} - \underline{\underline{T}}^{(E)} \right) \underline{U}_s^{(B)}$, $\underline{\underline{G}}^{(E,2)} = \underline{\underline{T}}^{-(E)} \underline{\underline{U}}^{(E)}$.

Equations (10) together with the set of equations

$$\underline{\underline{T}}^{(k)} \underline{Q}^{(k)} = \underline{\underline{T}}^{+(k)} \underline{Q}^{(k+1)} + \underline{\underline{T}}^{-(k)} \underline{Q}^{(k-1)}, \quad k = 3, \ldots, N-2$$

(11)

form a closed system.

In the case of homogeneous waveguide $\underline{\underline{T}}^{\pm(B)} = \underline{\underline{T}}^{\pm(E)} = \underline{\underline{T}}^{\pm(k)} = \underline{\underline{T}}^+ = \underline{\underline{T}}^-$ and $\underline{\underline{T}}^{(B)} = \underline{\underline{T}}^{(E)} = \underline{\underline{T}}^{(k)} = \underline{\underline{T}}$ we can find from (10) and (11) that $R_s^{(B)} = 0$ and $\underline{Q}^{(k)} = \left( \lambda_{s_0}^{(+)} \right)^{k-1} V_{s_0} \underline{U}_{s_0}$.

## 2. NUMERICAL RESULTS

In the new CASCIE-M code we have changed the boundary conditions (2) to the boundary conditions (10) described above. The difference between CASCIE and CASCIE-M codes is shown in Figure 1. In the new code we used a standard procedure EVCCG from IMSL MATH Fortran Library for calculating of matrix eigen vectors and eigen values which are needed for using the open boundary conditions.

As an example of using CASCIE-M, field distributions were calculated in a non-uniform accelerating structure based on a structured (disk-loaded) waveguide. Non-uniform accelerating structures have a specific feature. They are designed in such a way that they have the required regular field distribution only at a fixed working frequency within the propagation band. As for other frequencies, the field distribution has not been studied.

In addition, the question arises about the possibility of generating left-running waves in inhomogeneous waveguides and media [9,10,11,12,13,14]. Using the open boundary conditions without additional reflections from couplers can clarify this issue.

To check the correctness of the boundary conditions (10) and the procedure for their implementation, calculations of the reflection $R$ and transmission vectors $V$ were carried out for several homogeneous disk-loaded waveguides. Obtained results showed that the developed procedure really realizes the open boundary conditions. For example, for homogeneous disk-loaded waveguide with cells that coincide with the first cell of the CERN structure (see Table 1) the modules of entries of vectors $R$ [3] and $V$ are equal to: $|R_1| = 1.1\text{E-6}, |R_2| = 2.7\text{E-15}, |R_3| = 1.2\text{E-18}, |R_4| = 8.9\text{E-19}$; $|V_1| = 0.64, |V_2| = 6.8\text{E-11}, |V_3| = 5.8\text{E-14}, |V_4| = 1.0\text{E-16}$. In all cases studied, the reflection coefficient in the passband was small $|R_1| < 1.\text{E-5}$.

The accelerator section developed at CERN [15] was chosen as a prototype for the structured waveguide under consideration. The main characteristics of a model section are presented in Table 1.

The structured waveguide, which was investigated using the CASCIE-M code, consists of $N_H$ ($N_H \geq 4$) uniform cells that coincide with the first structure cell, then 26 regular cells of the CERN structure and finally $N_H$ uniform cells that coincide with the last structure cell. The dispersion curves of homogeneous waveguides are shown in Figure 2. It can be seen that there are frequency intervals where the incident wave propagates in the first homogeneous part and does not propagate in the last part.

---

[3] In all calculations we limited ourselves to four terms of the expansion of the longitudinal electric field (see [6]).

For tuning the nonuniform structured waveguide we used the phase method (see [9]).

Table 1 Structure Parameters

|  |  | CERN [15] |
|---|---|---|
| Frequency, GHz | 11.994 | 11.994 |
| RF phase advance per cell, rad. | $2\pi/3$ | $2\pi/3$ |
| Input, Output iris radii ($b_{k,1}$), mm | 3.15, 2.35 | 3.15, 2.35 |
| Input, Output iris thickness ($d_{k,1}$), mm | 1.67, 1.00 | 1.67, 1.00 |
| Structure period $D = d_{k,1} + d_{k,2}$, mm | 8.3317 | 8.3317 |
| Input, Output group velocity, % of c | 1.44, 0.76 | 1.44, 0.76 |
| First and last cell Q-factor | 5560, 5560 | 5536, 5738 |
| Number of regular cells | 26 | 26 |

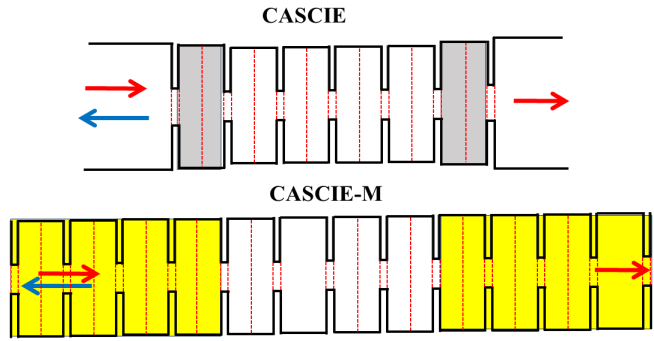

**Figure 1** Schematic representation of the structures that can be studied based on CASCIE and CASCIE-M codes

The CASCIE-M code makes it possible to calculate the field distribution at different frequencies without taking into account the influence of reflection from the couplers. In Figure 3 the distributions of complex longitudinal electric field $E_z(r = 0, z)$ along the z axis of 26 cells are presented.

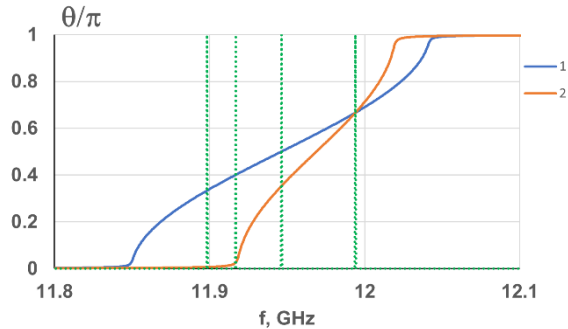

Figure 2 Dispersive characteristics of homogeneous disk-loaded waveguides: 1- cell sizes equal sizes of the first cell of the CERN structure, 2- cell sizes equal sizes of the last cell of the CERN structure

The first two pictures (1a, 2a) correspond the case when the wave travels through 26 cells, the last two pictures – when the wave reflects from turning point.

As noted above, in the passband the reflection coefficient from a homogeneous waveguide is practically zero. For an inhomogeneous waveguide, reflection occurs. The dependence of the reflection coefficient on frequency (see Figure 4) differs from the similar dependence for a waveguide with couplers. It does not have a regular sequence of minimum and maximum values associated with reflections from the couplers and the formation of resonance conditions. This indicates that the reflections are of a different nature and are associated with inhomogeneity. The value of the reflection coefficient at a frequency $f = 11.994$ GHz (working frequency) $R_1 \approx 8E-3$ practically coincide with the reflection coefficient value of the section with the two couples $R_1 \approx 7.9E-3$ [9].

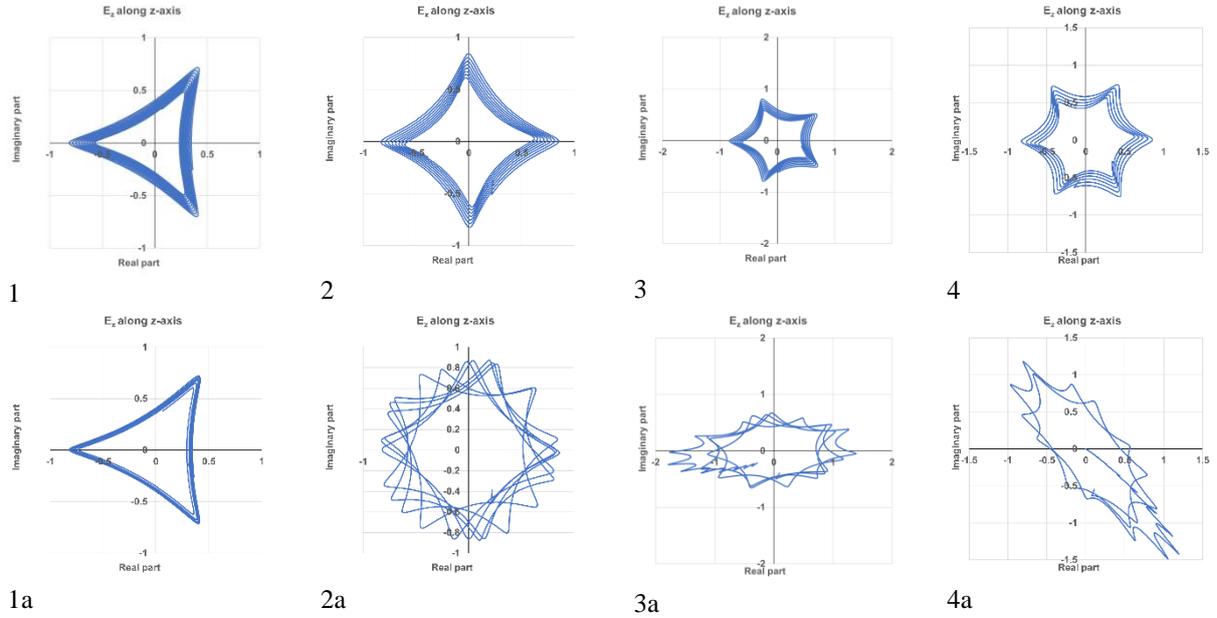

Figure 3 Complex longitudinal electric field $E_z(r=0,z)$ of 26 cells along the z axis: 1,2,3,4- homogeneous cells, which sizes equal sizes of the first cell of the CERN structure, 1- $\theta = 2\pi/3$, 2- $\theta = \pi/2$, 3- $\theta = 2\pi/5$, 4- $\theta = \pi/3$; 1a,2a,3a,4a - inhomogeneous cells, which sizes equal sizes of the cell of the CERN structure; 1,1a – f=11.994GHz, 2,2a- f=11.9464GHz, 3,3a- f=11.9168GHz, 4,4a – f=11.8984 GHz (see Figure 2).

$$\tag{12}$$

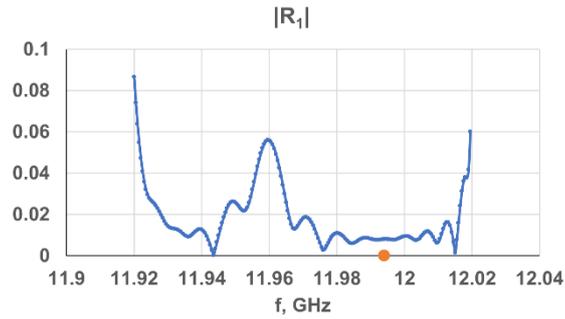

Figure 4 Dependence of reflection coefficient on frequency

Electromagnetic fields in a non-periodic structured waveguide with the ideal metal walls can be represented in the form of such series

$$\vec{H}(\vec{r}) = \sum_{s=-\infty}^{s=\infty} C_s(z)\vec{H}_s^{(e,z)}(\vec{r}), \tag{13}$$

$$\vec{E}(\vec{r}) = \sum_{s=-\infty}^{s=\infty} C_s(z)\vec{E}_s^{(e,z)}(\vec{r}), \tag{14}$$

where $\vec{E}_s^{(e,z)}(\vec{r}), \vec{H}_s^{(e,z)}(\vec{r})$ are modified eigen vector functions obtained by generalizing the eigen $\vec{E}_s^{(e)}, \vec{H}_s^{(e)}$ vectors of a homogeneous waveguide by special continuation of the geometric parameters [7,8].

The most useful is the case when we can use only the first terms in (13) and (14) (a single-mode representation)

$$\vec{E}_1(\vec{r}) = \vec{E}_1^{(+)}(\vec{r}) + \vec{E}_1^{(-)}(\vec{r}) = C_1(z)\vec{E}_1^{(e,z)}(\vec{r}) + C_{-1}(z)\vec{E}_{-1}^{(e,z)}(\vec{r}). \tag{15}$$

Representing the total field as the sum of known functions with unknown scalar coefficients, a system of coupled equations that determines the dependence of coefficients $C_1(z), C_{-1}(z)$ on the longitudinal coordinate can be obtained [7,8]. To use this system it is necessary to know the accuracy of expansion (15).

We can check out the accuracy of (15) by calculating the relative error of the representation of electric field

$$\chi = \left|\frac{E_z(0,z) - E_{1,z}(0,z)}{E_z(0,z)}\right| \tag{16}$$

We can also check the accuracy of the positive single-mode representation by computing the corresponding error

$$\chi^{(+)} = \left| \frac{E_z(0,z) - E_{1,z}^{(+)}(0,z)}{E_z(0,z)} \right| \qquad (17)$$

In (16) and (17) $E_z(0,z)$ is the longitudinal component of electric field calculated using the CASCIE-M code, $C_1(z)$ and $C_{-1}(z)$ were calculated by direct expansion the fields $\vec{E}(\vec{r})$ and $\vec{H}(\vec{r})$, calculated using the CASCIE-M code, in terms of vector functions $\vec{E}_s^{(e,z)}(\vec{r}), \vec{H}_s^{(e,z)}(\vec{r})$ [8].

Figure 5 and Figure 6 show the relative errors in representing the electric field for different frequencies. The error in representing the electric field by a single mode (15) is small. The relative error in representing the electric field by a positive single mode is almost two orders of magnitude greater. This indicates the important role of this component.

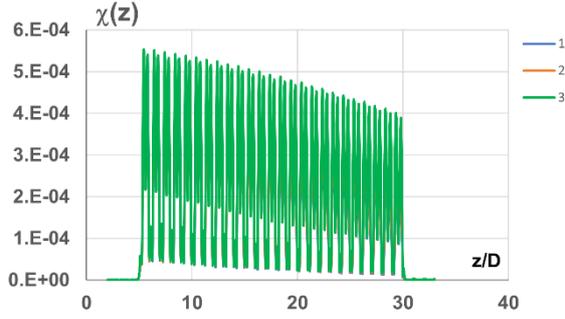
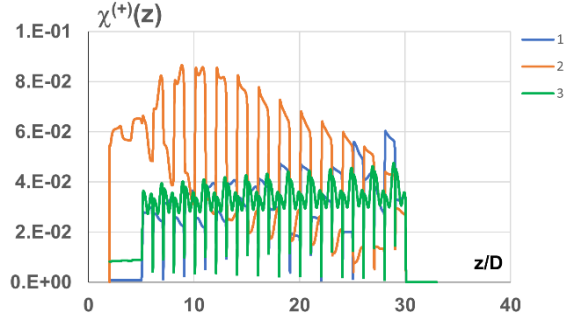

Figure 5 Relative error of representing the electric field by a single mode: 1 - f=11.9436 GHz (min reflection), 2 – f=11.96 GHz (max reflection), 3 – f=11.994 GHz (working frequency)

Figure 6 Relative error of representing the electric field by a positive single mode: 1 - f=11.9436 GHz (min reflection), 2 – f=11.96 GHz (max reflection), 3 – f=11.994 GHz (working frequency)

The component $\vec{E}_1^{(+)}(\vec{r})$ is the modified right travelling wave. Pictures 1a-3b of Figure **7** confirm it. As for the $\vec{E}_1^{(-)}(\vec{r})$ component, its physical meaning can be unambiguously determined only for a homogeneous waveguide. For inhomogeneous waveguide this component has a complex spatial dependence (see 1c-3d). Phase dependencies can be either decreasing on average (1d,2d) or increasing (3d).

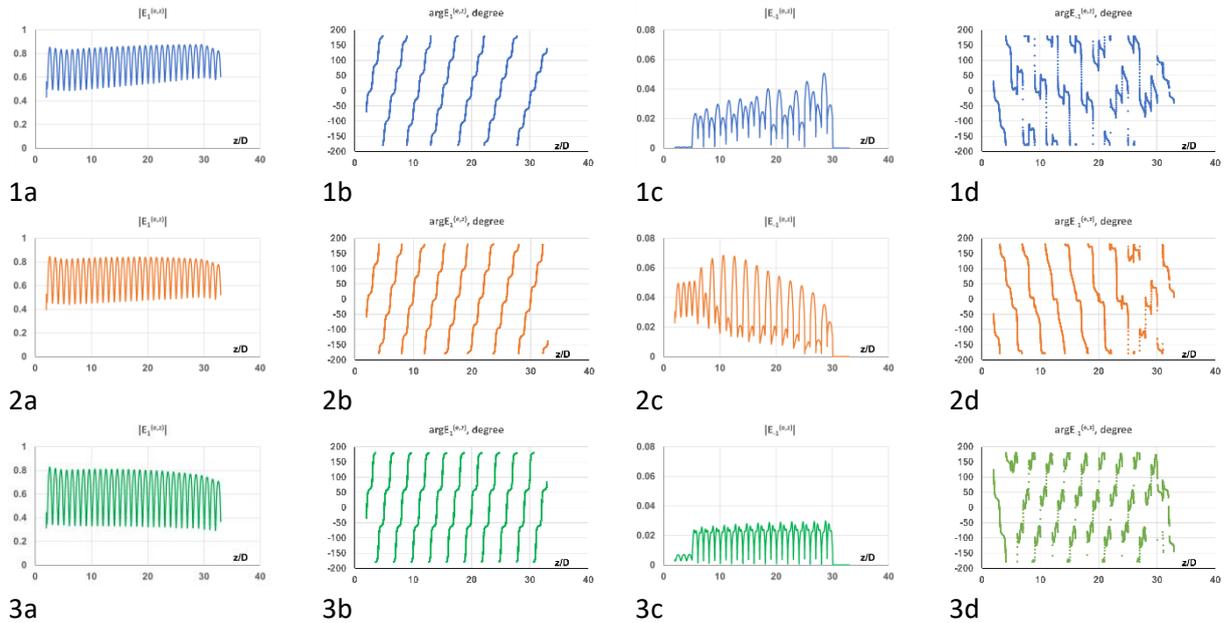

Figure 7 Dependencies of amplitudes (1a-3a, 1c-3c) and phases (1b-3b, 1d-3d) of field components $E_{1,z}^{(+)}(r=0,z)$ and $E_{1,z}^{(-)}(r=0,z)$ on longitudinal coordinate for different frequencies: 1 - f=11.9436 GHz (min reflection), 2 – f=11.96 GHz (max reflection), 3 – f=11.994 GHz (working frequency)

The above results show that the use of the open boundary conditions in the coupled integral equations method gives new possibility in the study of the characteristics of structured waveguides.

## CONCLUSIONS

Results of testing of the modify CASCIE code aimed at implementing open boundary conditions are presented. The accelerator section developed at CERN was chosen as a prototype for the structured waveguide under testing. Results of testing the CASCIE-M code confirms that the implementation of matrix open boundary conditions gives possibility to consider the structure in which waves enter and exit without additional reflections from couplers.

It was shown that the dependence of the reflection coefficient on frequency differs from the similar dependence for a waveguide with couplers. It does not have a regular sequence of minimum and maximum values associated with reflections from the couplers and the formation of resonance conditions. This indicates that the reflections are of a different nature and are associated with inhomogeneity. The proposed modification of the coupled integral equation method allows us to investigate the accuracy of the field expansion on which coupled mode theory can be constructed that describes structured waveguides.

Coupled mode theory that describes structured waveguides is based on the field expansion in terms of modified eigen vector functions obtained by generalizing the eigen vectors of a homogeneous waveguide by special continuation of the geometric parameters. Calculations show that there are waveguide geometries when single-mode approach gives good results. Under single-mode representation the electromagnetic field is the sum of two components: the first component $\vec{E}_1^{(+)}(\vec{r})$ is the modified right travelling wave. As for the second component $\vec{E}_1^{(-)}(\vec{r})$, its physical meaning is not defined completely. In previous works it was shown that this component can be a right travelling wave with a complex structure [9,10,14]. Results presented in this work show that for inhomogeneous waveguide this component can also has a decreasing phase distribution (similar to left traveling waves). Structure of this component is important for studying the interaction of the electron beam with the electromagnetic field. But it should be kept in mind that, in contrast to the homogeneous case, these two components are not the solution of Maxwell equations separately.